\documentclass[11pt,epsf,letterpaper]{article}%
\usepackage{color}
\usepackage{amsmath}
\usepackage{amsfonts}
\usepackage{verbatim}
\usepackage{amssymb}
\usepackage{graphicx}
\usepackage{epstopdf}
\usepackage{mathrsfs}%
\providecommand{\U}[1]{\protect\rule{.1in}{.1in}}
\begin{document}

\date{}
\title{Exact Hairy Black Holes and their Modification to the Universal Law of Gravitation}
\author{$^{(1)}$Andr\'{e}s Anabal\'{o}n and $^{(2)}$Julio Oliva.\\\textit{$^{(1)}$Departamento de Ciencias, Facultad de Artes Liberales y}\\\textit{Facultad de Ingenier\'{\i}a y Ciencias, Universidad Adolfo
Ib\'{a}\~{n}ez, Vi\~{n}a del Mar, Chile.}\\\textit{$^{(2)}$Instituto de Ciencias F\'{\i}sicas y Matem\'{a}ticas,
Universidad Austral de Chile, Valdivia, Chile.}}
\maketitle

\begin{abstract}
In this paper two things are done. First, it is pointed out the existence of
exact asymptotically flat, spherically symmetric black holes when a self
interacting, minimally coupled scalar field is the source of the Einstein
equations in four dimensions. The scalar field potential is the recently found
to be compatible with the hairy generalization of the Plebanski-Demianski
solution of general relativity. This paper describes the spherically symmetric
solutions that smoothly connect the Schwarzschild black hole with its hairy
counterpart. The geometry and scalar field are everywhere regular except at
the usual Schwarzschild like singularity inside the black hole. The scalar
field energy momentum tensor satisfies the null energy condition in the static
region of the spacetime. The first law holds when the parameters of the scalar
field potential are fixed under thermodynamical variation. Secondly, it is
shown that an extra, dimensionless parameter, present in the hairy solution,
allows to modify the gravitational field of a spherically symmetric black hole
in a remarkable way. When the dimensionless parameter is increased, the scalar
field generates a flat gravitational potential, that however asymptotically
matches the Schwarzschild gravitational field. Finally, it is shown that a
positive cosmological constant can render the scalar field potential convex if
the parameters are within a specific rank.

\end{abstract}

\section{Introduction.}

It is already more than forty years since Wheeler's original conjecture that
black holes have no hair. Although there was an intensive analysis on this
claim (for a review and references see \cite{Beke}), the situation, for the
minimally coupled scalar field, was clarified in the nineties. When the
spacetime is asymptotically flat there are two theorems that strongly
constraint the existence of black holes when the energy momentum satisfies the
dominant \cite{Heusler:1994wa} and the weak energy conditions
\cite{Sudarsky:1995zg}. When just the null energy condition is required there
is numerical evidence on the existence of asymptotically flat four dimensional
black holes \cite{Nucamendi:1995ex}.

It is particularly timely to consider the astrophyical relevance of this
problem. Actually, it has been pointed out that the angular and quadrupolar
momentum, $J$ and $Q$ respectively, of the black hole located at the center of
our galaxy, SgrA*, can be determined by the orbital precession of stars very
near to it, therefore allowing to check the relation $Q=-\frac{J^{2}}{Mc}$
that follows from the Kerr solution \cite{Sadeghian:2011ub}. Due to the
uniqueness and no hair theorems of asymptotically flat, four dimensional,
general relativity this would test the experimental validity of the hypothesis
that they involve.

In an attempt to construct a hairy rotating black hole, an exact family of
stationary and axisymmetric Petrov type D spacetimes was found when either a
(non)-minimally coupled scalar field \ or a generic non-linear sigma model is
the source of the Einstein equations \cite{Anabalon:2012ta}. It was found that
the most general potential that allows for the integration of the field
equations can be off-shell found if the geodesic motion on the scalar manifold
is integrable. The form of the spacetime metric is however independent of the
scalar manifold metric. This paper provides a discussion on some physical
implications that follow from the existence of these hairy black holes.

The simplest case of a single, minimally coupled, scalar field is very
interesting. Since the Lagrangian has no continuos symmetry, is not possible
to associate it with a conserved current, and therefore is a natural candidate
to be a dark matter component. The only source for the scalar field can be its
self interaction. The analysis of \cite{Anabalon:2012ta} allows to find, in
this case, a generic potential compatible with the Einstein equations:%

\begin{align}
V(\phi)  &  =\frac{\alpha}{\nu^{2}\kappa}\left(  \frac{\nu-1}{\nu+2}%
\sinh(\left(  1+\nu\right)  \phi l_{\nu})+\frac{\nu+1}{\nu-2}\sinh(\left(
1-\nu\right)  \phi l_{\nu})+4\frac{\nu^{2}-1}{\nu^{2}-4}\sinh\left(  \phi
l_{\nu}\right)  \right) \nonumber\\
&  +\frac{\Lambda\left(  \nu^{2}-4\right)  }{6\kappa\nu^{2}}\left(  \frac
{\nu-1}{\nu+2}e^{-\left(  \nu+1\right)  \phi l_{\nu}}+\frac{\nu+1}{\nu
-2}e^{\left(  \nu-1\right)  \phi l_{\nu}}+4\frac{\nu^{2}-1}{\nu^{2}-4}e^{-\phi
l_{\nu}}\right)  . \label{pot1}%
\end{align}
This is the most general potential allowing for uncharged, stationary and
axisymmetric, Petrov type D solutions of the Plebanski-Demianski type. Here
$\kappa=8\pi G$ where $G$ is the Newton constant$,$ $l_{\nu}=\left(
\frac{2\kappa}{\nu^{2}-1}\right)  ^{\frac{1}{2}},$ $\Lambda$ is the
cosmological constant and $\alpha$, $\nu$ are parameters of the potential.
This potential still makes sense when $\Lambda=0$ and, as is shown in this
work, the spherically symmetric solutions of the Einstein equation are
continuos deformations of the Schwarzschild black hole. Therefore, an
interesting result of this paper is to single out the first exact, uncharged,
asymptotically flat, black hole with everywhere regular geometry and matter
field except at the usual Schwarzschild singularity.

When $\Lambda=0$, the existence of these modifications to the physically
relevant Schwarzschild spacetime are pertinent to the motion of test particles
following geodesics in this geometry. It is important to stress that no
difference is expected to arise in the gravitational field outside of a star
since, in these cases, a discontinuity appears in the derivative of the
gravitational field at the surface of the star, and the lack of a conserved
current for the scalar field make this disconuity incompatible with its
existence in the first place. The modified gravitational field introduced and
discussed in this work is relevant as a model for black holes. In this case,
the dimensionless parameter $\nu$, plays a very important role in setting the
strenght of the gravitational field. Actually, for large enough values of the
parameter $\nu$,$\,$the strenght of the gravitational field of a hairy black
hole can be made esentially flat all the way from its surface up to regions as
far from the location of the event horizon as the model would require.
Moreover, the fact that the parameter $\nu$ does not enter in the Komar mass
allows to introduce an extra parameter in the gravitational field of a black
hole, providing a new astrophyiscal tool to fit the measured gravitational
field of any black hole configuration to this exact, analytical model, derived
from general relativity. The existence of scalar fields has been already
considered to be relevant to stellar kinematics \cite{AmaroSeoane:2010qx}, and
it has been noted that these models will be experimentally tested in the
future gravitational wave measurements \cite{AmaroSeoane:2012tx}.

Most of the paper is restricted to the case when $\Lambda=0$. In this case the
scalar field potential is unbounded from below, which, would suggest that the
solution is unstable. However, this is just an artifact of the $\Lambda=0$
limit and, as it is explicitly shown, for $\Lambda>0$ the potential can be
made convex sitting at its global minimum at infinity.

To close this introduction we would like to acknowledge that the fact that it
is possible relaxing the boundary conditions for gravitating scalar fields in
anti de Sitter spacetime \cite{Henneaux:2004zi} is what fuelled the
expectation that exact solutions of this system should exist
\cite{Martinez:2002ru}. However, it turns out that the black holes of
\cite{Anabalon:2012ta} still make sense when the cosmological constant vanishes.

The outline of the paper is as follows: in the second section the solutions
and the potential are described and the energy momentum tensor is shown to
satisfy the null energy condition. The third section describes the existence
of two kinds of solutions in this hairy black hole family. The fourth section
is devoted to the computation of the Komar mass and how it satisfies the first
law of black hole thermodynamics. A further degeneration in the configuration
is shown to exist. The fifth section describes the behavior of the
gravitational potential that a geodesic test particle feels in this
background, the dimensionless parameter $\nu$ sets the strenght of it. The
last section describes how a positive cosmological constant improves the
convexity of the potential. Finally, some remarks are made on the results of
this paper.

The notation follows~\cite{wald}. The conventions of curvature tensors are
such that a sphere in an orthonormal frame has positive Riemann tensor and
scalar curvature. The metric signature is taken to be $(-,+,+,+)$. Greek
letters are in the coordinate tangent space. Since we set $8\pi G=\kappa$ and
$c=1=\hbar$, the gravitational constant has units of lenght squated $\left[
\kappa\right]  =L^{2}.$

\section{The exact hairy black holes.}

As discussed in the introduction, this paper study the first exact,
asymptotically flat, solutions of the classical model:%

\begin{equation}
S(g,\phi)=\int d^{4}x\sqrt{-g}\left[  \frac{R}{2\kappa}-\frac{1}{2}g^{\mu\nu
}\partial_{\mu}\phi\partial_{\nu}\phi-V(\phi)\right]  ,
\end{equation}
with field equations:%

\begin{equation}
R_{\mu\nu}-\frac{1}{2}g_{\mu\nu}R=\kappa T_{\mu\nu},
\end{equation}%
\begin{equation}
T_{\mu\nu}=\partial_{\mu}\phi\partial_{\nu}\phi-\frac{1}{2}g_{\mu\nu}%
\partial_{\alpha}\phi\partial_{\beta}\phi g^{\alpha\beta}-g_{\mu\nu}V(\phi),
\end{equation}

\begin{equation}
V(\phi)=\frac{\alpha}{\nu^{2}\kappa}\left(  \frac{\nu-1}{\nu+2}\sinh(\left(
1+\nu\right)  \phi l_{\nu})+\frac{\nu+1}{\nu-2}\sinh(\left(  1-\nu\right)
\phi l_{\nu})+4\frac{\nu^{2}-1}{\nu^{2}-4}\sinh\left(  \phi l_{\nu}\right)
\right)  . \label{potnoL}%
\end{equation}
The following configurations are exact solution of this system
\cite{Anabalon:2012ta}:%

\begin{equation}
ds^{2}=\Omega(r)(-F(r)dt^{2}+\frac{dr^{2}}{F(r)}+d\Sigma^{2}), \label{M1}%
\end{equation}%
\begin{equation}
\Omega(r)=\frac{\nu^{2}\eta^{\nu-1}r^{\nu-1}}{\left(  r^{\nu}-\eta^{\nu
}\right)  ^{2}},
\end{equation}

\begin{equation}
F(r)=\frac{r^{2-\nu}\eta^{-\nu}\left(  r^{\nu}-\eta^{\nu}\right)  ^{2}}%
{\nu^{2}}+\left(  \frac{1}{\left(  \nu^{2}-4\right)  }-\left(  1+\frac
{\eta^{\nu}r^{-\nu}}{\nu-2}-\frac{\eta^{-\nu}r^{\nu}}{\nu+2}\right)
\frac{r^{2}}{\eta^{2}\nu^{2}}\right)  \alpha, \label{M3}%
\end{equation}

\begin{equation}
\phi=l_{\nu}^{-1}\ln(r\eta^{-1}),
\end{equation}
where $l_{\nu}=\left(  \frac{2\kappa}{\nu^{2}-1}\right)  ^{\frac{1}{2}}$ and
$d\Sigma^{2}$ is the line element of a unit two-sphere. $\eta$ is the only
integration constant of the black hole. The solution and theory are invariant
under the transformation $\nu\rightarrow-\nu$.

The energy momentum of the scalar field, in a comoving tetrad, has the form
$T^{ab}=diag(\rho,p_{1},p_{2},p_{2})$ and, in the static regions of the
spacetime, defined by $F(r)>0$, satisfies the null energy condition:
\begin{equation}
\rho+p_{2}=0,\qquad\rho+p_{1}=\frac{\left(  \nu^{2}-1\right)  \left(  r^{\nu
}-\eta^{\nu}\right)  ^{2}F(r)}{2r\nu^{2}\eta^{\nu-1}r^{\nu}}>0.
\end{equation}
In the hairless limit, $\nu=1$, the change of coordinates $r=\eta-\frac{1}{y}$
brings the hairy solution (\ref{M1})-(\ref{M3}) to the familiar Schwarzschild
black hole:%

\begin{equation}
ds^{2}=-(1-\frac{2M}{y})dt^{2}+\frac{dy^{2}}{1-\frac{2M}{y}}+y^{2}d\Sigma.
\label{K}%
\end{equation}
where $M=\frac{3\eta^{2}+\alpha}{6\eta^{3}}$.

The asymptotic behavior of the metric functions at $r=\eta$ is:%
\begin{equation}
\Omega(r)=\frac{1}{(r-\eta)^{2}}+\frac{1-\nu^{2}}{12\eta^{2}}-\frac{1-\nu^{2}%
}{12\eta^{3}}(r-\eta)+O(r-\eta)^{2},
\end{equation}

\begin{equation}
F(r)=(r-\eta)^{2}+\frac{3\eta^{2}+\alpha}{3\eta^{3}}(r-\eta)^{3}+O(r-\eta
)^{4}.
\end{equation}
It follows from these expressions that the leading behavior of the metric is
the same than the Schwarzschild solution with the radial coordinate given by
$\rho=\eta-\frac{1}{r}$. The cases with $\nu=2$ and $\nu=\infty$ are special,
and will be treated in a forthcoming publication where an exhaustive
classification of the solutions for this potential will be made.

\section{The two branches.}

One can notice that the potential (\ref{potnoL}) has a different behavior at
$\phi=\pm\infty$. From this observation it follows that there are two
solutions, depending on the branch that one is considering. For further
analysis is better to parametrize the solution with the dimensionless
coordinate $x=\frac{r}{\eta}$, such that now the asymptotic region is at
$x=1$. The convention of this paper is such that the solution:%

\begin{equation}
ds^{2}=\Omega(x)(-F(x)dt^{2}+\frac{\eta^{2}dx^{2}}{F(x)}+d\Sigma^{2}),
\end{equation}%
\[
\Omega(x)=\frac{\nu^{2}x^{\nu-1}}{\eta^{2}\left(  x^{\nu}-1\right)  ^{2}},
\]%
\begin{equation}
F(x)=\frac{x^{2-\nu}\left(  x^{\nu}-1\right)  ^{2}\eta^{2}}{\nu^{2}}+\left(
\frac{1}{\left(  \nu^{2}-4\right)  }-\left(  1+\frac{x^{-\nu}}{\nu-2}%
-\frac{x^{\nu}}{\nu+2}\right)  \frac{x^{2}}{\nu^{2}}\right)  \alpha,
\end{equation}%
\begin{equation}
\phi=l_{\nu}^{-1}\ln(x),
\end{equation}
with $x>1$ is defined as the positive branch while the negative branch is the
one defined for $x<1$.

\section{The Komar mass and the degeneration of the configurations.}

The computation of the Komar mass is straightforward. The result is given by:%

\begin{equation}
M_{-}=\frac{3\eta^{2}+\alpha}{6\eta^{3}G}. \label{mass}%
\end{equation}
The subscript ($-$) indicates that this is the mass for the region $x<1$. The
change in the orientation of the outward normal implies that the positive
branch has mass $M_{+}=-M_{-}$. The integration constant of the problem is
$\eta$ and the parameters of the Lagrangian are $\alpha$ and $\nu$. Given
these two parameters there are two configurations, the positive and the
negative branch. Therefore, given the two parameters, the boundary conditions
and the mass, each of the branches is completely characterized after solving
$\eta$ from:%

\begin{equation}
3\mu\eta^{3}+3\eta^{2}+\alpha=0, \label{cubi}%
\end{equation}
where $\mu=2M_{+}G$. It is clear that (\ref{cubi}) always has one real
solution. The discriminant of this cubic is:%
\begin{equation}
-27\alpha\left(  4+9\mu^{2}\alpha\right)  .
\end{equation}
The degeneration in $\eta$-s for each branch can be described as follows:

\begin{itemize}
\item \thinspace$\alpha=-\frac{4}{9\mu^{2}}$. In this case there is a double
degeneration on each branch given by the values $\eta=\frac{1}{3\mu}$ and
$\eta=-\frac{2}{3}\frac{1}{\mu}$.

\item $\alpha<-\frac{4}{9\mu^{2}}<0$. In this case there is only one $\eta$.

\item $-\frac{4}{9\mu^{2}}<\alpha<0$. There is a triple degeneration.

\item $\alpha>0$. There is only one black hole.
\end{itemize}

The case with $\alpha=0$ is pathological when $\Lambda=0$, and excluded from
this analysis. This degeneration of the solutions is very interesting from a
thermodynamical point of view, which although out of the scope of this work,
is presented here for completeness on the description of the black holes.

\subsection{The entropy and the first law.}

The entropy is one quarter of the area:%

\begin{equation}
S=\frac{A}{4G}=\frac{\pi\Omega(r_{+})}{G}=\frac{\pi}{G}\frac{\nu^{2}x_{+}%
^{\nu-1}}{\eta^{2}\left(  x_{+}^{\nu}-1\right)  ^{2}},
\end{equation}
where $x_{+}=\frac{r_{+}}{\eta}$ is the solution to $F(x_{+})=0.$ The
temperature is defined to make the Euclidean continuation smooth:%

\begin{equation}
T_{\pm}=\pm\frac{\eta}{4\pi}\frac{x_{+}\left(  1- x_{+}^{\nu}\right)  (\left(
\nu-2\right)  \left(  1+x_{+}^{\nu+2}\right)  +\left(  \nu+2\right)  \left(
x_{+}^{\nu}-x_{+}^{2}\right)  )}{\left(  x_{+}^{\nu+2}\left(  \nu
^{2}-4\right)  -\nu^{2}x_{+}^{\nu}+\left(  \nu+2\right)  x_{+}^{2}+\left(
2-\nu\right)  x_{+}^{2\nu+2}\right)  }.
\end{equation}
where the subscript $\pm$ indicates the temperatures for the black holes
defined by the positive and negative branch respectively. With these results
at hand it is straightforward to check that:%

\begin{equation}
\delta M_{K}=T\delta S,
\end{equation}
where, the variation let the parameters of the Lagrangian, $\alpha$ and $\nu$,
fixed. All these quantities have smooth limits when $\nu=2$.

\section{The gravitational field.}

A test particle moving in this gravitational field satisfies the geodesic equation:%

\begin{equation}
\left(  \Omega(r)\dot{r}\right)  ^{2}+F(r)L^{2}+\Omega(r)F(r)=E^{2},
\end{equation}
where $L=\Omega(r)\dot{\phi}$ and $E=\Omega(r)F(r)\dot{t}$ are conserved
quantities and the dot stands for the derivative respect to the geodesic
affine parameter. If the coordinate $\dot{\rho}=\Omega(r)\dot{r}$ is
introduced it is clear that the $\frac{\Omega(r)F(r)-1}{2}$ term is the actual
gravitational potential that a test particle feels on this background.

Therefore, to undertand the departure of the geodesic motion from the expected
from the Schwarzschild black hole ($\nu=1$), is enough to compare minus the
lapse function, $-g_{tt}=\Omega(x)F(x)$, for differents values of $\nu$. It is
imporant to remark that all the corrections of the geodesic motion in the
Schwarzschild geometry to the Newtonian behavior are proportional to the
angular momentum of the particle. The lapse function in the $\nu=1$ case is
just the Keplerian potential.

\subsection{Different gravitational fields for the same black hole area.}

To model the same black hole in the different theories, defined by the
different values of $\alpha$ and $\nu$, it is necessary to fix its area. Thus,
let us fix it as follows:%

\begin{equation}
\Omega(x_{+},\eta,\nu)=\rho_{+}^{2}\Longrightarrow\eta=\frac{\left\vert
\nu\right\vert x_{+}^{\frac{\nu-1}{2}}}{\rho_{+}\left\vert x_{+}^{\nu
}-1\right\vert },
\end{equation}
where $\rho_{+}$ defines the area of the horizon to be $4\pi\rho_{+}^{2}$ and
$x_{+}$ is the location of the horizon defined by $F(x_{+})=0$. This defintion
of $\eta$ allows to writte $\Omega$ as:%

\begin{equation}
\Omega(x,x_{+},\rho,\nu)=\rho_{+}^{2}\frac{x^{\nu-1}\left(  x_{+}^{\nu
}-1\right)  ^{2}}{\left(  x^{\nu}-1\right)  ^{2}x_{+}^{\nu-1}}. \label{Area}%
\end{equation}

\begin{figure}[h]
\centering
\includegraphics[height=2.8in,width=3.8in]{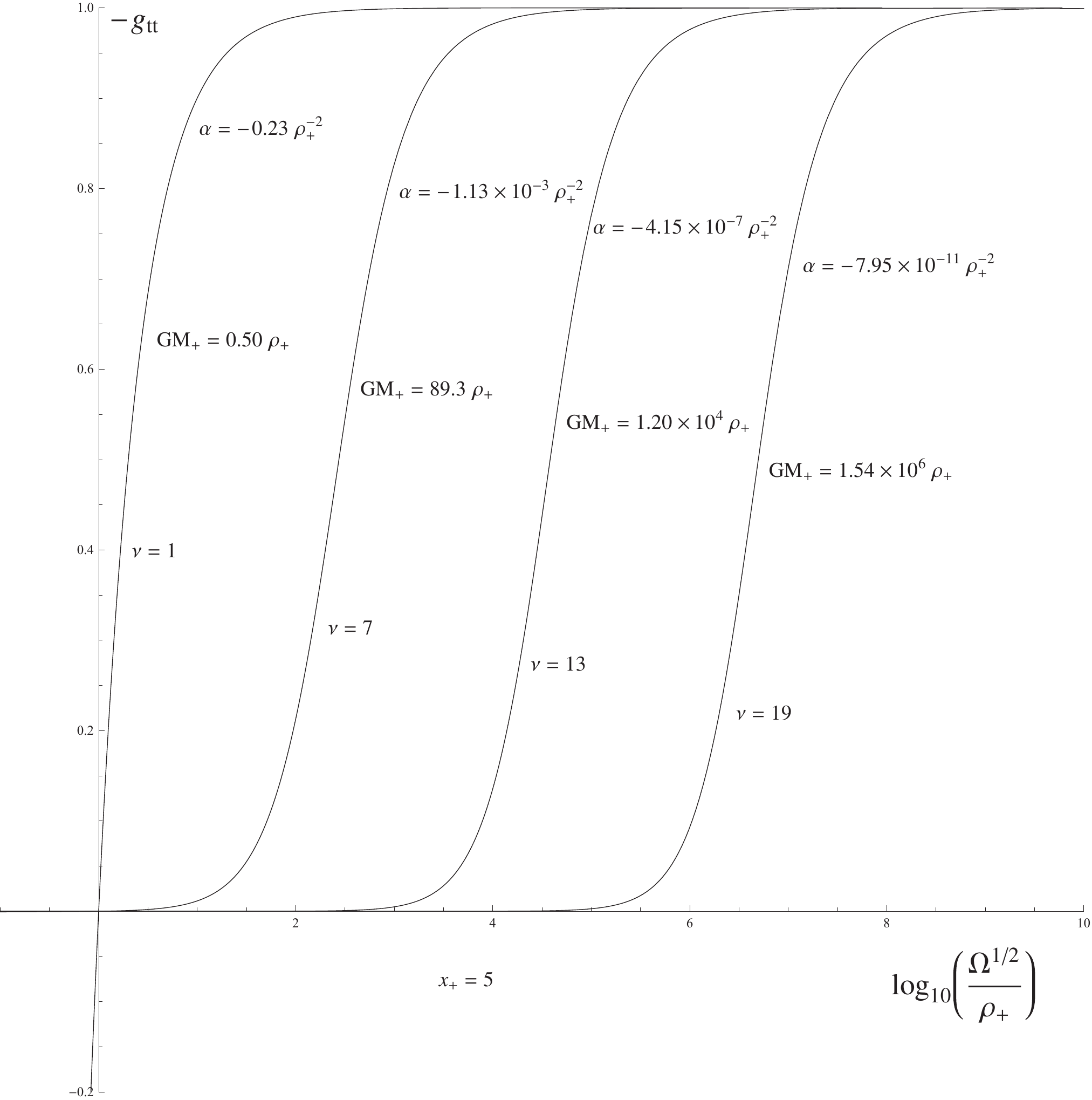}\caption{The positive branch.
$\Omega(x)F(x)$ for a fixed horizon area. The horizontal axis is $\log
_{10}(\sqrt{\Omega(x)}/\rho_{+}).$ This function is the analogue of the
logarithm of the radial coordinate in the Schwarzschild geometry,
\textbf{normalized to be zero at the horizon}.}%
\label{ramamascontodo}%
\end{figure}

The definition of the location of the horizon can be used to find the value of
the parameter $\alpha$ in terms of $x_{+}$ and $\eta$ and $\nu$. Finally,
$\eta$ and $\alpha$ can be replaced in the negative of the lapse function to
obtain $F(x)\Omega(x)$ as a function of $x$, $x_{+}$ and $\nu$. These graphs
are independent of $\rho_{+}$. They depend on $x_{+}$ through the relation of
the Komar mass and $\rho_{+}$ and the relation of $\alpha$ and $\rho_{+}^{-2}$.

\begin{figure}[h]
\centering
\includegraphics[height=2.8in,width=3.8in]{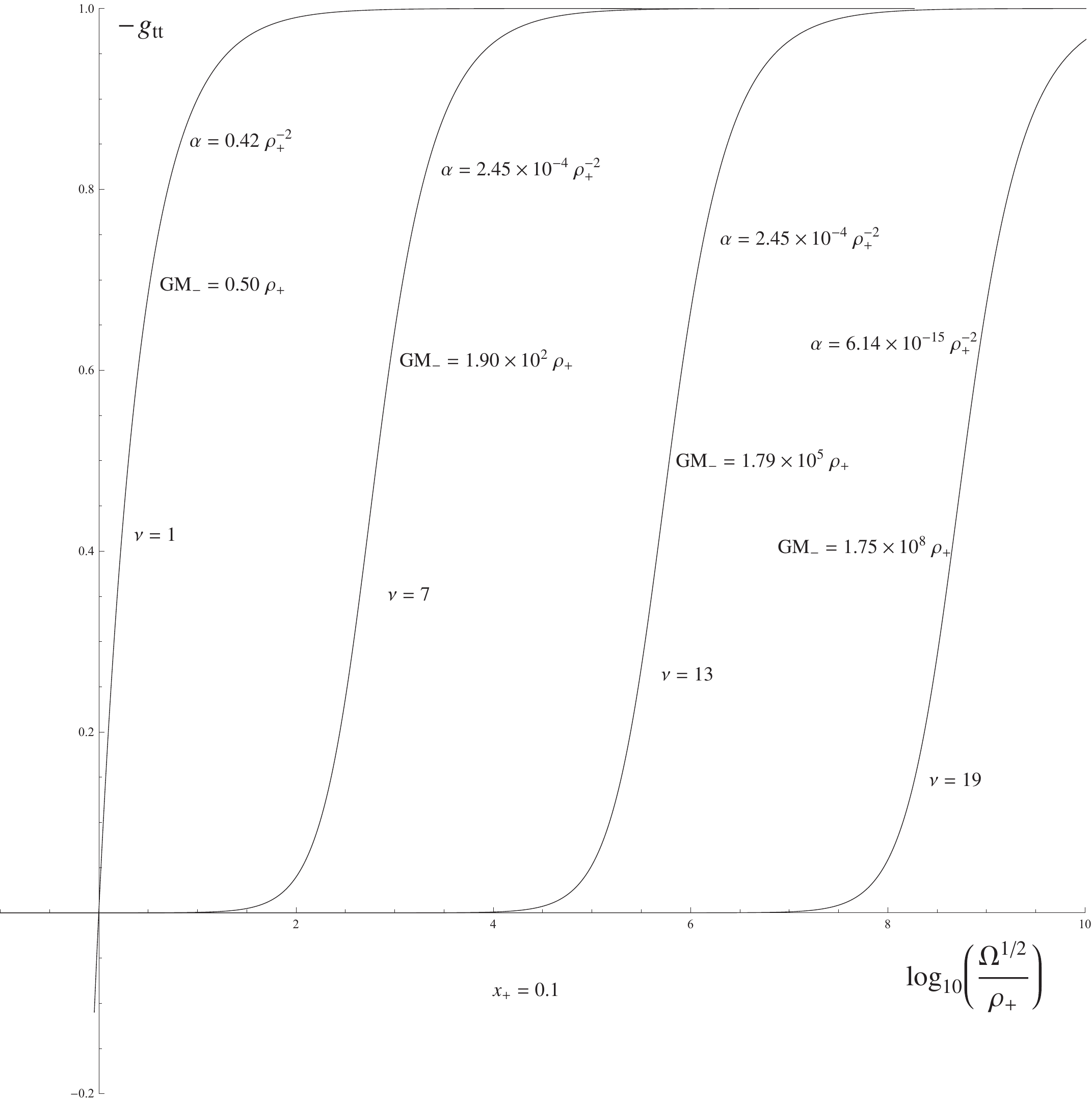}\caption{The negative
branch. $\Omega(x)F(x)$ for a fixed horizon. The horizontal axis is $\log
_{10}(\sqrt{\Omega(x)}/\rho_{+}).$This function is the analogue of the
logarithm of the radial coordinate in the Schwarzschild geometry,
\textbf{normalized to be zero at the horizon}.}%
\label{ramamenoscontodo}%
\end{figure}The fact that the horizontal axis is on a logarithmic scale shows
how flat can the gravitational field be made in the presence of the scalar
field. This effect is a purely general relativistic effect. It is very
interesting to note that the second derivative of the scalar field potential
is zero at $\phi=0$, and therefore the scalar field is massless at infinity.
However this massless scalar, due to the non-linear structure of the Einstein
equations, has a strong effect in the gravitational field; it smoothly enhance
the gravitational field from the Schwarzschild solution to large deviations
from it.

\section{The scalar field potential.}

When $\Lambda=0$ the potential is odd, namely $V^{O}(\phi)=-V^{O}(-\phi)$. It
follows that the stability of these solutions is an issue that should be
addressed. While the actual calculation on the mechanical stability is a
non-trival task, that will be addresed on a future work, this section shows
that there are positive values of the cosmological constant that make the
potential convex.

The more general potential (\ref{pot1}), modifies the solution presented here
in a simple way:%

\begin{equation}
F^{\Lambda}(r)=F(r)-\frac{\Lambda}{3}.
\end{equation}

To understand the structure of the potential when $\Lambda>0$, we note that,
as follows from (\ref{pot1}), the potential is even if $\alpha=\alpha
_{E}\equiv\frac{\Lambda\left(  \nu^{2}-4\right)  }{6}$. In this case
(\ref{pot1}) takes the following form:%

\begin{equation}
V^{\alpha=\alpha_{E}}(\phi)=\frac{\Lambda\left(  \nu^{2}-4\right)  }%
{6\kappa\nu^{2}}\left(  \frac{\nu-1}{\nu+2}\cosh(\left(  1+\nu\right)  \phi
l_{\nu})+\frac{\nu+1}{\nu-2}\cosh(\left(  1-\nu\right)  \phi l_{\nu}%
)+4\frac{\nu^{2}-1}{\nu^{2}-4}\cosh\left(  \phi l_{\nu}\right)  \right)  .
\end{equation}

From the form of the even potential it follows that its leading asymptotic
form is:%

\begin{equation}
V^{\alpha=\alpha_{E}}(\phi=\pm\infty)=\frac{\Lambda\left(  \nu-2\right)
\left(  \nu-1\right)  }{12\kappa\nu^{2}}\exp((\nu+1)\left\vert \phi\right\vert
l_{\nu}).
\end{equation}

More generally, the potential has the following asymptotic expansions for
large values of $\phi$:%

\begin{align}
V(\phi &  =+\infty)=\frac{\alpha}{2\nu^{2}\kappa}\left(  \frac{\nu-1}{\nu
+2}\right)  \exp(\left(  1+\nu\right)  \phi l_{\nu}),\\
V(\phi &  =-\infty)=\frac{1}{2\nu^{2}\kappa}\left(  \frac{\nu-1}{\nu
+2}\right)  \left(  -\alpha+\frac{\Lambda\left(  \nu^{2}-4\right)  }%
{3}\right)  \exp(-\left(  1+\nu\right)  \phi l_{\nu}). \label{pot2}%
\end{align}

When $\alpha=0$ the leading behavior of the potential is:%

\begin{equation}
V^{\alpha=0}(\phi=+\infty)=\frac{\Lambda\left(  \nu+2\right)  \left(
\nu+1\right)  }{6\kappa\nu^{2}}e^{\left(  \nu-1\right)  \phi l_{\nu}},\qquad
V^{\alpha=0}(\phi=-\infty)=\frac{\Lambda\left(  \nu-2\right)  \left(
\nu-1\right)  }{6\kappa\nu^{2}}e^{-\left(  \nu+1\right)  \phi l_{\nu}}.
\end{equation}

It can be noticed from (\ref{pot2}) that when $\alpha=2\alpha_{E}%
=\frac{\Lambda\left(  \nu^{2}-4\right)  }{3}$ the leading behavior at infinity
is also different. In fact, $V^{\alpha=0}(\phi)=V^{\alpha=2\alpha_{E}}(-\phi
)$. It is possible to make a classification of all the possible behaviors of
the potential for different values of $\Lambda$, $\nu$ and $\alpha$. This
classification will be reported separatedly because it contains more than
twenty different situations. The case that is more compelling in favor of the
stability is:

\textbf{When} $\nu>2$\textbf{ and }$\frac{\Lambda\left(  \nu^{2}-4\right)
}{3}>\alpha>0$\textbf{. The potential is positive for large values of
}$\left\vert \phi\right\vert $\textbf{. The scalar field potential has a
minimum at }$\phi=0$.

The reparametrization of the solution in terms of $x_{+}$, $\rho_{+}$ and
$\nu$ implies that:%

\begin{equation}
\alpha=\frac{(\nu^{2}-4)\nu^{2}x_{+}^{\nu}(\Lambda\rho_{+}^{2}-3x_{+})}%
{3\rho_{+}^{2}\left(  x_{+}^{\nu}\left(  x_{+}^{\nu+2}\left(  \nu-2\right)
+x_{+}^{2}\left(  4-\nu^{2}\right)  +\nu^{2}\right)  -x_{+}^{2}(\nu+2)\right)
}.
\end{equation}
It follows from this expression that all the cases of the negative branch of
the previous section can satisfy $\frac{\Lambda\left(  \nu^{2}-4\right)  }%
{3}>\alpha>0$ for a range of positive values for $\Lambda$. These values can
be made as small as required for small enough $x_{+}$.

\section{Final Remarks.}

Before this work there was only one uncharged, spherically symmetric black
hole in four dimensional general relativity; now there is an infinite family
of them. Inded, we make reference to configurations that are everywhere
regular except at the usual Schwarzschild singularity.

As follows from the analysis of the lapse function, the scalar field allows to
have much larger than Schwarzschild ammounts of mass in a given region of the
spacetime. If this scalar field would actually exist, relatively small black
holes can be much more massive than one would expect based on the
Schwarzschild solution. These smaller but massive black holes strongly modify
the universal law of gravity not only in its near surroundings but also in
arbitrarily far regions from their location. It is tempting to conjecture that
these geometries can be usefull to describe the geodesic motion of actual test
particles in an astrophysical situation, like the well known flat galactic
rotation curves. However, this can be considered as a serious posibility only
after a study of the stability of these solutions is made. This, indeed, will
be addressed in a future publication.

\section{Acknowledgments.}

We thank the useful coments of Nicolas Grandi about this work. Research of
A.A. is supported in part by the Alexander von Humboldt foundation and by the
Conicyt grant Anillo ACT-91: \textquotedblleft Southern Theoretical Physics
Laboratory\textquotedblright\ (STPLab). The work of J. O. was supported by
FONDECYT grant 11090281.



\begin{thebibliography}{99}                                                                                               %


\bibitem {Beke}J.~D.~Bekenstein, \textit{`No Hair': Twenty--five Years After},
chapter in \textit{ Proceedings of the Second International Andrei D. Sakharov
Conference in Physics\/}, edited by I.~M.~Dremin and A.~M.~Semikhatov (World
Scientific, Singapore, 1997).

\bibitem {Heusler:1994wa}M.~Heusler, \textquotedblleft A Mass bound for
spherically symmetric black hole space-times,\textquotedblright%
\ Class.\ Quant.\ Grav.\ \textbf{12} (1995) 779 [gr-qc/9411054].


\bibitem {Sudarsky:1995zg}D.~Sudarsky, ``A Simple proof of a no hair theorem
in Einstein Higgs theory,,'' Class.\ Quant.\ Grav.\ \textbf{12} (1995) 579.


\bibitem {Nucamendi:1995ex}U.~Nucamendi and M.~Salgado, \textquotedblleft
Scalar hairy black holes and solitons in asymptotically flat
space-times,\textquotedblright\ Phys.\ Rev.\ D \textbf{68} (2003) 044026
[gr-qc/0301062].


\bibitem {Sadeghian:2011ub}L.~Sadeghian and C.~M.~Will, \textquotedblleft
Testing the black hole no-hair theorem at the galactic center: Perturbing
effects of stars in the surrounding cluster,\textquotedblright%
\ Class.\ Quant.\ Grav.\ \textbf{28} (2011) 225029 [arXiv:1106.5056 [gr-qc]].
C.~M.~Will, \textquotedblleft Testing the general relativistic no-hair
theorems using the Galactic center black hole SgrA*,\textquotedblright%
\ arXiv:0711.1677 [astro-ph].


\bibitem {Anabalon:2012ta}A.~Anabalon, ``Exact Black Holes and Universality in
the Backreaction of non-linear Sigma Models with a potential in (A)dS4,''
arXiv:1204.2720 [hep-th].


\bibitem {AmaroSeoane:2010qx}P.~Amaro-Seoane, J.~Barranco, A.~Bernal and
L.~Rezzolla, ``Constraining scalar fields with stellar kinematics and
collisional dark matter,'' JCAP \textbf{1011} (2010) 002 [arXiv:1009.0019
[astro-ph.CO]].


\bibitem {AmaroSeoane:2012tx}P.~Amaro-Seoane, ``Stellar dynamics and
extreme-mass ratio inspirals,'' arXiv:1205.5240 [astro-ph.CO].


\bibitem {Henneaux:2004zi}M.~Henneaux, C.~Martinez, R.~Troncoso and
J.~Zanelli, ``Asymptotically anti-de Sitter spacetimes and scalar fields with
a logarithmic branch,'' Phys.\ Rev.\ D \textbf{70} (2004) 044034
[hep-th/0404236].
M.~Henneaux, C.~Martinez, R.~Troncoso and J.~Zanelli, ``Asymptotic behavior
and Hamiltonian analysis of anti-de Sitter gravity coupled to scalar fields,''
Annals Phys.\ \textbf{322} (2007) 824 [hep-th/0603185].


\bibitem {Martinez:2002ru}C.~Mart\'{\i}nez, R.~Troncoso, and J.~Zanelli,
\textquotedblleft De Sitter black hole with a conformally coupled scalar field
in four dimensions,\textquotedblright\ Phys.\ Rev.\ D \textbf{67}, 024008
(2003) [arXiv:hep-th/0205319].
C.~Mart\'{\i}nez, J.~P.~Staforelli, and R.~Troncoso, \textquotedblleft Charged
topological black hole with a conformally coupled scalar
field,\textquotedblright\ Phys.\ Rev.\ D \textbf{74}, 044028 (2006)
[arXiv:hep-th/0512022].
E.~Radu and E.~Winstanley, \textquotedblleft Conformally coupled scalar
solitons and black holes with negative cosmological
constant,\textquotedblright\ Phys.\ Rev.\ D \textbf{72}, 024017 (2005)
[arXiv:gr-qc/0503095].
A.~Anabalon, H.~Maeda, \textquotedblleft New Charged Black Holes with
Conformal Scalar Hair,\textquotedblright\ Phys.\ Rev.\ \textbf{D81 } (2010)
041501.
[arXiv:0907.0219 [hep-th]]. C.~Charmousis, T.~Kolyvaris, and
E.~Papantonopoulos, \textquotedblleft Charged C-metric with conformally
coupled scalar field,\textquotedblright\ Class. Quant. Grav. \textbf{26},
175012 (2009) [arXiv:0906.5568 [gr-qc]].
T.~Kolyvaris, G.~Koutsoumbas, E.~Papantonopoulos and G.~Siopsis,
\textquotedblleft A New Class of Exact Hairy Black Hole
Solutions,\textquotedblright\ Gen.\ Rel.\ Grav.\ \textbf{43} (2011) 163
[arXiv:0911.1711 [hep-th]].
M.~J.~Duff, J.~T.~Liu, \textquotedblleft Anti-de Sitter black holes in gauged
N = 8 supergravity,\textquotedblright\ Nucl.\ Phys.\ \textbf{B554 } (1999)
237-253. [hep-th/9901149]. T.~Kolyvaris, G.~Koutsoumbas, E.~Papantonopoulos
and G.~Siopsis, \textquotedblleft Einstein Hair,\textquotedblright%
\ arXiv:1111.0263 [gr-qc].
S.~G.~Saenz and C.~Martinez, \textquotedblleft Anti-de Sitter massless scalar
field spacetimes in arbitrary dimensions,\textquotedblright\ arXiv:1203.4776
[hep-th].
A.~Anabalon, F.~Canfora, A.~Giacomini and J.~Oliva, \textquotedblleft Black
Holes with Primary Hair in gauged N=8 Supergravity,\textquotedblright%
\ arXiv:1203.6627 [hep-th].
A.~Anabalon and A.~Cisterna, \textquotedblleft Asymptotically (anti) de Sitter
Black Holes and Wormholes with a Self Interacting Scalar Field in Four
Dimensions,\textquotedblright\ arXiv:1201.2008 [hep-th].
A.~Anabalon and H.~Maeda, \textquotedblleft New Charged Black Holes with
Conformal Scalar Hair,\textquotedblright\ Phys.\ Rev.\ D \textbf{81} (2010)
041501 [arXiv:0907.0219 [hep-th]].
A.~Anabalon, F.~Canfora, A.~Giacomini and J.~Oliva, ``Black Holes with Primary
Hair in gauged N=8 Supergravity,'' arXiv:1203.6627 [hep-th].
Y.~Bardoux, M.~M.~Caldarelli and C.~Charmousis, ``Conformally coupled scalar
black holes admit a flat horizon due to axionic charge,'' arXiv:1205.4025
[hep-th].


\bibitem {wald}R.~M.~Wald, \textit{General Relativity}, (University of Chicago
Press, 1984), p.491.
\end{thebibliography}
\end{document}